\documentclass[twocolumn,prb,amsmath,amssymb,floatfix,superscriptaddress,showpacs]{revtex4}
\usepackage{color}
\usepackage{graphicx}
\usepackage{dcolumn}
\usepackage{bm}
\usepackage{times}
\begin{document}


\def\QAF{${\bf Q}_{\rm AF}$}

\title{Low-energy phonons and superconductivity in
Sn$_{0.8}$In$_{0.2}$Te}

\author{Zhijun~Xu}
\affiliation{Condensed Matter Physics and Materials Science
Department, Brookhaven National Laboratory, Upton, New York 11973,
USA} \affiliation{Physics Department, University of California,
Berkeley, California 94720, USA} \affiliation{Materials Science
Division, Lawrence Berkeley National Laboratory, Berkeley,
California 94720, USA}
\author{J. A.~Schneeloch}
\affiliation{Condensed Matter Physics and Materials Science
Department, Brookhaven National Laboratory, Upton, New York 11973,
USA} \affiliation{Department of Physics, Stony Brook University,
Stony Brook, New York 11794, USA}
\author{Ruidan~Zhong}
\affiliation{Condensed Matter Physics and Materials Science
Department, Brookhaven National Laboratory, Upton, New York 11973,
USA} \affiliation{Materials Science and Engineering Department,
Stony Brook University, Stony Brook, New York 11794, USA}
\author{J. A. Rodriguez-Rivera}
\affiliation{NIST Center for Neutron Research, National Institute of
Standards and Technology, Gaithersburg, Maryland 20899, USA}
\affiliation{Department of Materials Science \&\ Engineering,
University of Maryland, College Park, MD 20742, USA}
\author{L.~Harriger}
\affiliation{NIST Center for Neutron Research, National Institute of
Standards and Technology, Gaithersburg, Maryland 20899, USA}
\author{R. J. Birgeneau}
\affiliation{Physics Department, University of California, Berkeley,
California 94720, USA} \affiliation{Materials Science Division,
Lawrence Berkeley National Laboratory, Berkeley, California 94720,
USA}
\author{Genda~Gu}
\author{J.~M.~Tranquada}
\author{Guangyong~Xu}
\affiliation{Condensed Matter Physics and Materials Science
Department, Brookhaven National Laboratory, Upton, New York 11973,
USA}

\date{\today}

\begin{abstract}
We present neutron scattering measurements on low-energy phonons
from a superconducting ($T_{c} = 2.7$~K) Sn$_{0.8}$In$_{0.2}$Te
single crystal sample. The longitudinal acoustic phonon mode and one
transverse acoustic branch have been mapped out around the (002)
Bragg peak for temperatures of 1.7 K and 4.2 K.  We observe a
substantial energy width of the transverse phonons at energies
comparable to twice the superconducting gap; however, there is no
change in this width between the superconducting and normal states.
We also confirm that the compound is well ordered, with no
indications of structural instability.
\end{abstract}


\maketitle

\section{Introduction}

The discovery of the topological insulators
(TIs),\cite{Fu2007,Hasan2010} which are insulating (theoretically)
in the bulk but have metallic surface states present due to their
topologically nontrivial electronic structure, have attracted great
scientific interest. These materials can be categorized by the
symmetry by which their surface states are protected. For example,
the "topological crystalline insulators"
(TCIs)\cite{Fu2011,Tanaka2012} have surface states protected by
certain crystal point group symmetries rather than by time-reversal
symmetry invariance (TRI-TIs),\cite{Fu2007,Xia2009} as observed in
other compounds.

The studies of TRI-TIs and TCIs have stimulated the search for an
even more exotic state of matter, the topological superconductor
(TS), whose surface states are predicted to be Majorana
fermions.\cite{Fu2008,Qi2011} One candidate is
Cu$_{x}$Bi$_{2}$Se$_{3}$, a superconductor arising from Cu-doping of
the TRI-TI Bi$_{2}$Se$_{3}$.\cite{Hor2010} The most important
evidence for (or against) the existence of TS in particular
compounds has come from the presence (or absence) of a zero-bias
conductance peak (ZBCP) in point-contact spectra. Such a peak may be
indicative of unconventional superconductivity, and calculations
have shown that every possible unconventional pairing symmetry for
Cu$_{x}$Bi$_{2}$Se$_{3}$ should be topologically nontrivial
\cite{fu_odd-parity_2010, Sasaki2011}. Some groups have reported
such a peak in
Cu$_{x}$Bi$_{2}$Se$_{3}$\cite{Sasaki2011,Kirzhner2012,Chen2012},
though other groups have raised doubts on these measurements
\cite{levy_experimental_2013,peng_absence_2013}. Furthermore, the
diamagnetic shielding fraction in Cu$_{x}$Bi$_{2}$Se$_{3}$ varies
significantly and is usually very low\cite{Hor2010}, limiting
possible studies.

Another system of interest is based on SnTe.  Pure SnTe has been
proposed\cite{Hsieh2012} and demonstrated\cite{Tanaka2012} to be a
topological crystalline insulator; furthermore, it has a
ferroelectric phase at low
temperature\cite{Pawley1966,Kobayashi1976} and intriguing
thermoelectric properties at high temperature.\cite{PbTe3} More
relevant to the present study is that Sn$_{1-\delta}$Te is
superconducting at temperature well below 1~K.\cite{Allen1969}
Substitution of a small amount of In for Sn drives the ferroelectric
instability to zero, while increasing the superconducting
$T_c$.\cite{Erickson2009}  More recent studies have shown that $T_c$
grows linearly with In concentration, reaching $\sim4.5$~K for
Sn$_{0.55}$In$_{0.45}$Te.\cite{BalPRB,Zhong2013}

The superconductivity in Sn$_{1-\delta}$Te was originally explained
by electron-phonon coupling involving the scattering of carriers
between equivalent conduction
band-valleys.\cite{Allen1969,Cohen1964} It has been noted more
recently that the hybridization between valence and conduction bands
that leads to band inversion and topological effects also leads to
enhanced van Hove singularities, and that this increased density of
states might be helpful to superconductivity.\cite{He2014}  When
$T_c$ was raised by substitution of In, Erickson {\it et
al.}\cite{Erickson2009} considered the possibility of pairing
enhancement via negative $U$ centers associated with In.  Sasaki
{\it et al.}\cite{Sasaki2012} observed a zero-bias conduction peak
in a point-contact measurement on Sn$_{1-x}$In$_x$Te with $x=0.45$,
arguing that this is evidence for an odd-parity superconducting
state.  In contrast, Saghir {\it et al.}\cite{Saghir2014} have
measured the temperature dependence of the magnetic penetration
depth by muon spin rotation spectroscopy and find that it is
consistent with an $s$-wave gap.  They also find evidence for strong
coupling, which may be necessary given the large magnitude of the
normal-state resistivity found in In-doped SnTe
crystals.\cite{Zhong2013}  Clearly, resolving the character of the
superconducting state in this system is of considerable current
interest.


In this paper, we report low-energy neutron scattering measurements
on a superconducting single-crystal of Sn$_{0.8}$In$_{0.2}$Te with
$T_c=2.7$~K.  Finding no evidence for any structural instability at
low temperature, we have characterized longitudinal and transverse
acoustic phonon dispersions in this sample below and above $T_c$.
The phonon dispersion relations  in the superconducting sample are
consistent with these results in the parent compound
SnTe.\cite{Cowley1969,PbTe3}
For a superconductor in which pairing is due to electron-phonon
coupling, one expects to see characteristic changes in the phonon
self energy for phonon energies comparable to $2\Delta$, where
$\Delta$ is the energy of the superconducting
gap.\cite{Shapiro1975,Allen1997,Weber2008,Allen1974,Allen1983}
We find instead a surprisingly large energy width for phonon energy
$\hbar\omega\sim2\Delta$, with no significant change across $T_c$.
We discuss possible causes of this large self energy.

\section{Experimental Details}

The single-crystal sample used in this experiment was grown by a
modified floating-zone method~\cite{Zhong2013,Zhong2014} at
Brookhaven National Laboratory. The mass is 22~g. The zero field
cooled (ZFC) susceptibilities and resistivity, measured with a
superconducting quantum interference device (SQUID) magnetometer,
are shown in Fig.~\ref{fig:1}(a), suggesting a bulk $T_c\approx
2.7$~K. The samples have also been characterized by x-ray
diffraction, indicating a rocksalt cubic structure.\cite{Zhong2013}

Neutron scattering experiments were carried out on the triple-axis
spectrometer SPINS and on the Multi-Axis Crystal Spectrometer (MACS)
\cite{Rodriguez2008} located at the NIST Center for Neutron Research
(NCNR). We used horizontal beam collimations of
Guide-$80'$-S-$80'$-$240'$ (S = sample) with a fixed final energy of
5 meV on SPINS together with a cooled Be filter after the sample to
reduce higher-order neutrons. We used a fixed final energy of 5.1
meV, with BeO filters after the sample and a PG filter before the
sample, and horizontal collimations of open-PG-open-S-$90'$-BeO-open
on MACS.

The inelastic neutron scattering measurements have been performed in
the $(HHL)$ scattering plane. The lattice constants for this sample
are $a = b =c \approx 6.33$~\AA. The data are described in
reciprocal lattice units (r.l.u.) of $(a^*, b^*, c^*) = (2\pi/a,
2\pi/b, 2\pi/c)$.

\section{Results and Discussions}

\begin{figure}[t]
\includegraphics[width=0.9\linewidth]{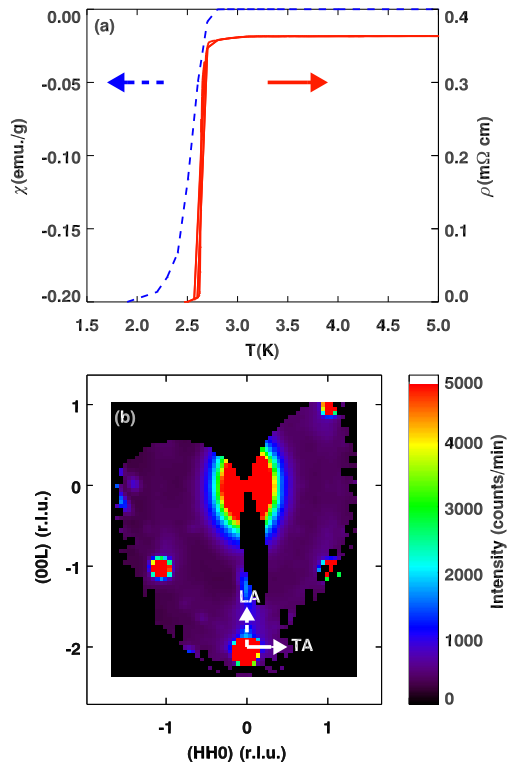}
\caption{(Color online) (a) Temperature dependence of ZFC
magnetization (blue dash line, left side) and resistivity (red solid
line, right side, from Ref.~\onlinecite{Zhong2013}). (b) Elastic
scans in scattering plane $(HHL)$ at 1.7~K. The dashed and solid
arrows show the longitudinal acoustic phonon (LA) and transverse
acoustic phonon (TA) measured around Bragg peak (002), respectively.
} \label{fig:1}
\end{figure}

The elastic scattering in the $(HHL)$ plane from MACS  at
$T=1.7~{\rm K} < T_c$ is shown in Fig.~\ref{fig:1}(b).  The
relatively low intensity of the (1,1,1) peak and the absence of
Bragg peaks at (1,1,0) or (0,0,1) are consistent with the rocksalt
cubic structure.~\cite{Zhong2013}  There is no evidence of any
superlattice or diffuse scattering that might indicate a structural
instability.
Hence, we focus our studies on the transverse (TA) and longitudinal
(LA) acoustic phonons around the strong (0,0,2) peak as shown in
Fig.~\ref{fig:1} (b).

\begin{figure}[t]
\includegraphics[width=0.9\linewidth]{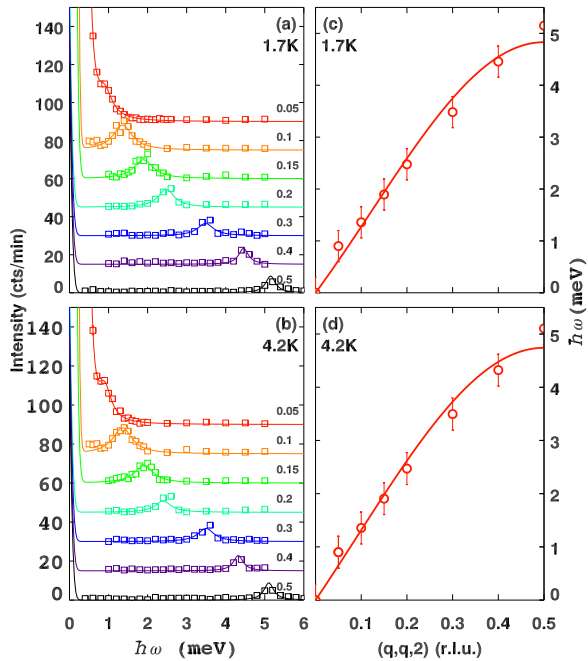}
\caption{(Color online) (a,b) Constant-$q$ cuts of phonon spectra
along transverse directions near (002) Bragg peak at (a) 1.7~K and
(b) 4.2~K and at various $\textbf{Q}$ positions with
$\textbf{Q}=(q,q,2)$, with $q=0.05$ (red), 0.1 (orange), 0.15
(green), 0.2 (teal), 0.3 (blue), 0.4 (purple), and 0.5 (black). The
data are taken on SPINS. (c,d) The TA$_1$ phonon dispersions around
(002) at (c) 1.7~K and (d) 4.2~K. The phonon energies are obtained
from fitting the energy cuts from (a)-(b) as described in the text.
The lines are guides to the eye. The errors are obtained from the
fitting process.} \label{fig:2}
\end{figure}

In Fig.~\ref{fig:2} (a)-(b), we plot constant-\textbf{Q} scans from
data taken at SPINS at various wave vectors along the transverse
direction to show the TA$_1$ phonon mode (polarized along [001] and
propagating along [110]). The phonon peaks are fitted with
Lorentzian functions, plus a Gaussian function describing the
elastic peak. The fitted curves are plotted with the data. We can
see that the TA phonons are all well defined. The energy center,
energy width, and intensity of the phonon modes can be obtained in
the fitting. The phonon dispersion relations are plotted in
fig.~\ref{fig:2} (c)-(d). The phonon dispersion relations measured
at both 1.7~K and 4.2~K are overall consistent with previous
inelastic neutron scattering measurements on the parent compound
SnTe.\cite{Cowley1969,PbTe3} Those studies reported that the band
top for the [110] acoustic phonon is around 5 meV, which is the same
as in our results.

\begin{figure}[t]
\includegraphics[width=0.9\linewidth]{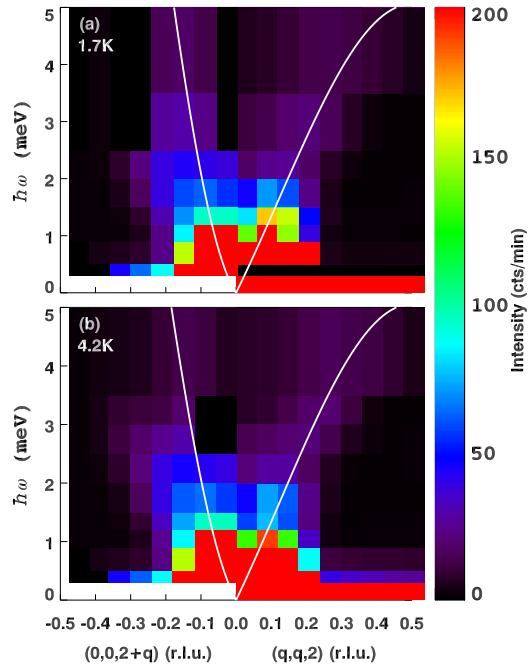}
\caption{(Color online) Phonon intensity plots in energy-momentum
space around Bragg peak (002) at (a) 1.7~K and (b) 4.2~K. For each
panel, the left half corresponds to longitudinal phonons going along
[001] direction; and the right half corresponds to transverse
phonons going along the [110] direction. The white lines are phonon
dispersions obtained from fitting of the energy cuts.} \label{fig:3}
\end{figure}

Mesh scans (constant energy measurements from MACS) were also
performed around the (002) Bragg peak  to map out the phonon modes
along both the transverse and longitudinal directions. We plot
intensity maps in energy-momentum space in Fig.~\ref{fig:3}. The
white lines shown here are guides to the eye that describe the
phonon dispersions. The dispersion relations of both the TA and LA
modes are in good agreement with previously reported data as well as
recent neutron scattering measurements on the parent compound
SnTe,\cite{Cowley1969,PbTe3} suggesting that a 20\% In doping does
not significantly modify the low-energy lattice dynamics in this
material.

In the parent SnTe and a similar compound PbTe, a soft transverse
optic (TO) mode has been observed, that condenses into a column type
intensity near zone-center at low
temperatures.~\cite{Pawley1966,Cowley1969,PbTe1,PbTe2,PbTe3,PbTe4}
Limited by the energy and $Q$ range of our measurements, we are not
able to map out the entire band for this TO mode. Nevertheless, we
did look for the column type intensity in our SPINS measurements
(Fig.~\ref{fig:1}), but have not noticed any additional intensity,
beyond the transverse acoustic phonons, that could suggest a soft TO
mode. This mode, associated with the ferroelectric instability, is
screened by free carriers\cite{Allen1969,Cowley1969}, so that it is
possible that the large In concentration in our sample has stiffened
the TO mode sufficiently that it is not present in our energy
window.  It is also possible that we have not detected it due to a
smaller TO phonon structure factor, compared to that of the acoustic
phonon, near (002).

\begin{figure}[t]
\includegraphics[width=0.9\linewidth]{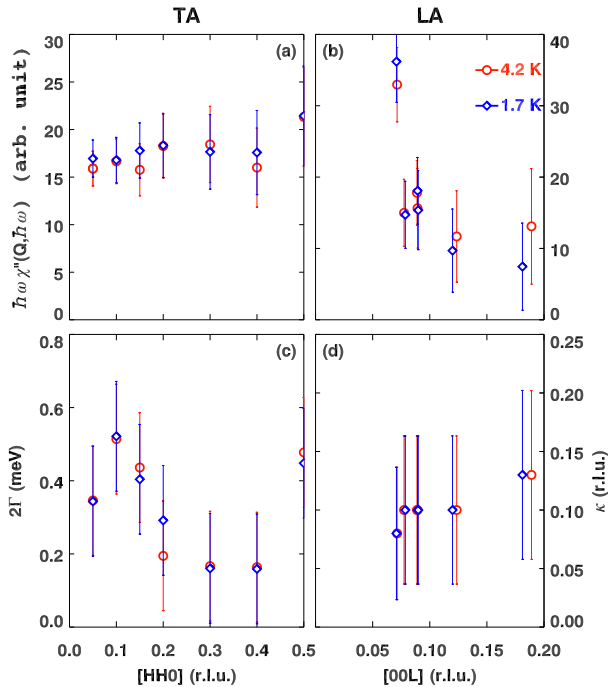}
\caption{(Color online) Summary of the fitting parameters. (a,b) The
dynamic response $\chi''$ multiplied by phonon energy $\hbar\omega$
for acoustic phonons measured along (a) transverse [110] direction,
and (b) longitudinal [001] directions at various q positions. (c)
The energy width $2\Gamma$ for the TA$_1$ phonon mode measured on
SPINS. (d) The $q$ width for the LA phonon measured on MACS. In
(a)-(b), the errors represent $\pm$ 1 standard deviation. In
(c)-(d), the errors are obtained from the fitting process.}
\label{fig:4}
\end{figure}

We now look at all the parameters obtained from the fits to the
data. The dynamic response $\chi''(\textbf{Q},\hbar\omega) =
(1-e^{-\hbar\omega/k_{B}T})S(\textbf{Q},\hbar\omega)$ is the
measured neutron scattering intensity divided by the Bose factor.
Here the natural temperature dependence of the phonon intensities
has been taken out so that we can look at any intrinsic changes
directly. The products $\hbar\omega\chi(Q,\omega)$ at different
temperatures ($T = 1.7$ and 4.2 K) are shown in the upper panels of
Fig.~\ref{fig:4}. This product remains almost constant for all $q$
values for the TA mode at both temperatures, suggesting a highly
harmonic mode. For the LA phonon mode, we see one high point for
$q=0.07$, while the product remains almost constant for all other
$q$ values. This high point at small $q$ could be a possible result
of the anomalously broad soft TO phonon mode near zone
center~\cite{PbTe3, PbTe4} observed in PbTe and SnTe at low
temperatures. The measured phonon energies and momentum widths are
shown in the lower panels of Fig.~\ref{fig:4}. The energy widths,
$2\Gamma$, which are inversely proportional to the phonon lifetime,
are obtained from the constant-Q scans from SPINS shown in
Fig.~\ref{fig:2}.
The $q$-widths, shown in Fig.~\ref{fig:4}~(d), are extracted from
the LA phonon measured on MACS.

Based on previous studies of the change in phonon self energy due to
superconducting order,\cite{Axe1973,Shapiro1975,Allen1997} we expect
to see the largest effects near the energy $2\Delta$, where $\Delta$
is the superconducting gap.  On cooling below $T_c$, one expects to
see an increase in the line width for $\hbar\omega\gtrsim 2\Delta$
and a decrease for $\hbar\omega\lesssim 2\Delta$  The recent work of
Saghir {\it et al.}\cite{Saghir2014} indicates strong coupling, with
$2\Delta\approx 4k_BT_c$, which yields an estimate of $2\Delta\sim
0.9$~meV for our crystal with $T_c=2.7$~K.   This is roughly
consistent with a point-contact measurement on a sample of
Sn$_{1-x}$In$_x$Te with $x=0.045$ $T_c=1.2$~K which observed
$2\Delta\sim 0.2$~meV.\cite{Sasaki2012}  The lowest phonon energy we
can measure is limited by the instrumental resolution ($\delta E
\sim$ 0.35 meV for SPINS, $\delta E\sim$ 0.5 meV for MACS at
$\hbar\omega$=0 meV) is around $\hbar\omega \alt 1$~meV, close to
the estimated value of $2\Delta$.

The effective energy resolution for the TA mode depends on how the
anisotropic resolution function matches up with the dispersion; this
is the focusing effect, which is discussed in Ref.
\onlinecite{Shapiro1975}.  For most of the wave vectors covered, the
velocity is relatively constant, and we calculate an effective
energy width of $\sim0.15$~meV, taking into account the sample
mosaic.  This is quite close to the measured widths for $H=0.3$ and
0.4, shown in Fig.~\ref{fig:4}(c), so those measurements appear to
be resolution limited.  The effective resolution width at the zone
boundary rises because of flattening of the dispersion.  In
contrast, the measured width near $H=0.1$ becomes much larger than
the effective resolution.  The phonon energy in this region is close
to $2\Delta$; however, we see no significant change in the energy
width between the superconducting and normal states.  While the
width changes due to superconducting order are expected to be
small,\cite{Shapiro1975}  the large, temperature-independent
contribution certainly makes their detection more challenging.


What could be the cause of this large energy width at small $q$?
One possibility is that is due to interactions with the soft TO
mode.  Another possibility is that these low-$q$ phonons are
involved in scattering electrons within a small pocket at the Fermi
surface.  According to Allen and Cohen,\cite{Allen1969} such
interactions contribute little to electron pairing; the main
contribution comes from scattering between different pockets, which
involves phonons at large $q$.   Further experimental work would be
necessary to identify such an effect in the phonon widths.

\section{Summary}

Overall, we have performed low energy neutron scattering
measurements on a single crystal of Sn$_{0.8}$In$_{0.2}$Te. There is
no evidence of structural instability in this sample. Two acoustic
phonon modes (TA$_1$ along [110] and LA along [001]) have been
mapped out at temperatures below and above $T_c$. The acoustic
phonon dispersion relations are consistent with those in the parent
topological insulator SnTe.  A large, temperature-independent phonon
energy width at $\hbar\omega\sim2\Delta$ obscures any change
associated with the onset of superconductivity, but indicates that
strong interactions may be present.


\acknowledgments

J.A.S. and R.Z. are supported by the Center for Emergent
Superconductivity, an Energy Frontier Research Consortium supported
by the Office of Basic Energy Science of the Department of Energy.
The work at Brookhaven National Laboratory and Lawrence Berkeley
National Laboratory was supported by the Office of Basic Energy
Sciences (BES), Division of Materials Science and Engineering, U.S.
Department of Energy (DOE), under Contract Nos.DE-AC02-98CH10886 and
DE-AC02-05CH1123, respectively. This work utilized facilities
supported in part by the National Science Foundation under Agreement
No. DMR-0944772.


\end{document}